\documentclass[12pt]{article}
\usepackage{epsfig}

\begin{document}

\begin{titlepage}

\begin{flushright}
\begin{tabular}{l}
                DESY 96-146 \\
                LNF-96/037(P) \\
                LBNL-39175\\
                hep-ph/9608213 \\
                July 1996   
\end{tabular}
\end{flushright}
\vspace{1.5cm}

\begin{center}
{\huge Charmed Mesons Fragmentation Functions} \\

\vspace{1.5cm}

{\large M. Cacciari$^a$, M. Greco$^{b}$, S. Rolli$^c$ and A. Tanzini$^{d}$} \\
\vspace{.5cm}

{\sl $^a$Deutsches Elektronen-Synchrotron DESY, Hamburg, Germany\\
\vspace{.2cm}
$^b$Dipartimento di Fisica E. Amaldi, Universit\`a di Roma III,\\ and INFN, 
Laboratori Nazionali di Frascati, Italy \\ 
\vspace{.2 cm}
$^c$INFN-Pavia, Italy and Lawrence Berkeley Laboratory, Berkeley, USA\\
\vspace{.2 cm}
$^d$Dipartimento di Fisica,
Universit\`a di Tor Vergata, Roma,\\
and INFN, Sezione di Roma2, Italy \\}
\end{center}

\vspace{1cm}

\begin{abstract}
Fragmentation functions for heavy-light mesons, like the charmed $D$, $D^*$
mesons, are proposed. They rest on next-to-leading QCD Perturbative
Fragmentation  Functions for heavy quarks, with the addition of  a
non-perturbative term describing  phenomenologically the quark $\to$ meson
transition. The cross section for production of large $p_T$ $D$, $D^*$ mesons
at the Tevatron is evaluated in this framework.
\end{abstract}

\vfill
\noindent\rule{7cm}{.2mm}
\small
\vspace{-.4cm}
\begin{tabbing}
e-mail addresses: \= cacciari@desy.de \\
                  \> greco@lnf.infn.it \\
		  \> rolli@fnal.gov \\
		  \> tanzini@vaxtov.roma2.infn.it
\end{tabbing}

\end{titlepage}

\section{Introduction}

Much experimental and theoretical work has been recently devoted to the study 
of heavy flavour production in hadronic collisions.
Theoretically, because the heavy quark mass is setting the scale in the 
perturbative expansion of QCD, acting as a cut-off for the infrared 
singularities, the most relevant features of this process are calculable 
within perturbation  theory. Indeed the calculation in perturbative QCD
of the differential and total cross sections to order $\alpha_s^3$ has been
performed \cite{NDE}, thus providing a firm basis for a detailed study of the 
properties of the bottom and charm quarks.

The NLO one particle inclusive differential distribution will however contain 
terms
of the kind  $\alpha_s\ln(p_T/m)$ which, in the large $p_T$ limit, will become 
large and will spoil the perturbative expansion of the cross section. This is 
reflected in a large sensitivity to the choice of the
renormalization/factorization scales and hence in a large uncertainty in the
theoretical prediction. In Ref.~\cite{cg1} this problem was tackled 
by introducing the technique of the Perturbative Fragmentation Functions (PFF)
through which these terms were resummed to all orders and the cross section was
shown to display a milder scale sensitivity.

When considering heavy meson inclusive production,
non perturbative effects are also quite important, especially for charmed mesons
(being $m_c\approx 1.5$ GeV), so they have to be estimated as better as possible
for a reliable calculation of the production cross sections.
They are normally introduced within the formalism of FF and indeed  a
next-to-leading order(NLO) analysis of FF into charmed mesons (in particular
$D,D^{*}$) including non-perturbative effects has been performed 
in Ref. \cite{cn} for $e^+ e^-$ annihilation processes up to LEP energies.
In this analysis however the charm component in the FF is
considered only, the other components giving a small contribution to the 
$e^+ e^-$ production cross section.

On the other hand, the contribution of gluon-gluon 
and quark-gluon scattering subprocesses to the production cross section
is relevant  in hadronic collisions, and the gluonic component 
can no longer be neglected.

Thus we proceed in this paper to the construction of a set of
NLO fragmentation functions for $D,D^{*}$ mesons, including
gluon, light and anticharm quark contribution. This set can therefore be used
in the calculation of large $p_T$ inclusive production cross section
for any hard collision process.

\section{Theoretical framework}

The general framework of  this analysis is the following: we'll consider 
the fragmentation into charm quark of any parton produced at large 
transverse momentum $p_T \gg m_c$, followed by the hadronization of the
charm quark into the  meson. Exploiting the difference in time scales
of the two processes, short for the perturbative fragmentation of the parton
into the charm and longer for the non-perturbative hadronization of the charm
in a meson, we can factor the overall fragmentation function of the parton $i$ 
into the meson $H$ in the following way:
\begin{equation}
D_i^H(z,\mu,m_c) = D_i^c(z,\mu,m_c) \otimes D_{np}^H(z)
\label{ansatz}
\end{equation}
the $\otimes$ symbol meaning the usual convolution operation. This expression
represents our ansatz for the fragmentation of a parton into a charmed meson.
The non perturbative part of the fragmentation is taken to be universal, i.e.
independent of the parton which produced the charm quark via perturbative
cascade. It is also independent of the scale at which the fragmentation
function is taken: all the evolution effects are dumped into the perturbative
part.

The first part of the process can be calculated with purely perturbative
techniques at an initial scale of the order of the charm mass, 
while we have to rely on phenomenological inputs to 
extract the hadronization non-perturbative effects at a fixed scale.
Then using the DGLAP 
evolution equations at NLO accuracy we can evaluate
fragmentation functions at the appropriate scale ${\cal O}(p_T)$, assuming 
that no scaling violation effects arise in the non perturbative part of the
fragmentation function.

The calculation of the perturbative part of the process has been carried
out in Ref. \cite{melenason}. For the reader's convenience we shall briefly 
report 
the main results of this analysis.
Using the factorization property, the charm quark production cross section
in $e^+ e^-$ collisions can be written as:
\begin{eqnarray}
&&{{d\sigma}\over{dx}}(x,Q,m_c)=\sum_i \int_x^1 {{d\hat \sigma_i}\over{dx}}
({x\over z},Q,\mu)D_i^c(z,\mu,m_c){dz\over z} \label{ee}
\end{eqnarray}
where $x$ is the energy fraction 
of the charm quark, $Q$ is the center-of-mass energy
and $m_c$ is the charm mass.
Eq.(1) shows that the cross section is factorized
into a short-distance term ${{d\hat \sigma_i}\over{dx}}$ 
for the production of the massless parton $i$, 
and  a parton FF $ D_i^c$
into the charm quark, evaluated at a scale $\mu$.
When $\mu$ is taken to be of the order of $m_c$, $D_i^c(x,\mu,m_c)$ 
is expressed in a perturbative expansion in powers of $\alpha_s$:
\begin{eqnarray}
&&D_i(z,\mu ,m_c)=d_i^{(0)}(z)+{{\alpha_s}\over {2\pi}}d_i^{(1)}(z,\mu ,m_c)
+O(\alpha_s^2) \label{DP}
\end{eqnarray}
Then using the perturbative expansion of the l.h.s. of eq.(\ref{ee}) 
one obtains the explicit expression of $d_i^{(0)}$ and $d^{(1)}_i$ coefficients.

This has been explicitly done is Ref. \cite{melenason}, obtaining the following
set of NLO initial conditions  in 
$\overline{MS}$ scheme for the fragmentation function of 
a charm quark, gluon and light quarks respectively, into  the 
charm quark:
\begin{eqnarray}
&&\hat D_c^c(x,\mu_0) = \delta(1-x) + {{\alpha_s(\mu_0) C_F}\over{2\pi}}\left[
{{1+x^2}\over{1-x}}\left(\log{{\mu_0^2}\over{m_c^2}} -2\log(1-x)
-1\right)\right]_+ \label{DQQ} \\ 
&&\hat D_g^c(x,\mu_0) = {{\alpha_s(\mu_0) T_f}\over{2\pi}}(x^2 + (1-x)^2)
\log{{\mu_0^2}\over{m_c^2}} \label{DgQ} \\
&&\hat D_{q,\bar q,\bar c}^c(x,\mu_0) = 0 \label{DqQ}
\end{eqnarray}
where $\mu_0$ is taken of the order of the charm quark mass, which we
fix in our analysis at 1.5 GeV.
As obvious, $D_g^c$ is of order $\alpha_s$,
and $D_{q,\bar q,\bar c}^c$ is 
zero in the NLO approximation, being of the order of $\alpha_s^2$.
Nonetheless, this component is generated at higher scales through 
the evolution with DGLAP equations, which involve a mixing of all
parton components of FF.

The PFF initial conditions (\ref{DQQ},\ref{DgQ},\ref{DqQ}), 
evolved up to the appropriate scale ${\cal O}(p_T)$ with NLO accuracy, can 
be used to evaluate the open charm production cross section
in the large $p_T$ region.
Indeed this allows the resummation of potentially large logarithms of the kind
$\alpha_s\log(p_T/m_c)$, arising from quasi-collinear
configurations, thus recovering a more reliable prediction of the 
$p_T$ spectrum at large transverse momentum
than the fixed order O($\alpha_s$) calculation.
This is discussed in detail in Ref. \cite{cg1} in the case of $b$ quarks and in
ref. \cite{cg2} for charm photoproduction.

In order to obtain the inclusive production of $D,D^{*}$ mesons,
one has to take into account further the hadronization of the
charm quark into the final charmed meson.
The Perturbative Fragmentation Functions (\ref{DQQ},\ref{DgQ},\ref{DqQ})
can be convoluted with a non-perturbative part, which we parametrize as
\begin{eqnarray}
&&D_{np}^H(z)=\langle n_H\rangle\, A (1-z)^{\alpha}z^{\beta} \label{nonpert}\\
&&{1\over A}=\int_0^1 (1-z)^{\alpha}z^{\beta}dz \nonumber
\end{eqnarray}
where the  parameters $\alpha$, $\beta$ and $\langle n_H\rangle$ have to be 
extracted from comparison with experimental data at a fixed scale.
Indeed $\alpha$ and $\beta$ have been obtained by Colangelo and Nason
in Ref. \cite{cn} by fitting  ARGUS data \cite{argus} 
for $D,D^{*}$ mesons fragmentation
functions in $e^+ e^-$ collisions at center-of-mass energy of 
$10.6$ GeV. They are reported in Table \ref{table1}. 
We'll use their determination, on the ground of our universality
assumption for the non-perturbative part of the fragmentation functions. It is
worth mentioning that we have challenged this universality assumption by
comparing our fragmentation functions (\ref{ansatz}), evolved up to 90 GeV and
using the ARGUS parameters
of Table \ref{table1},  with LEP data by OPAL \cite{opal}. Reasonable 
agreement has
been found, giving support to our hypothesis. It is also worth noting that
Colangelo and Nason original work \cite{cn} gives  a sub-set of our 
fragmentation functions only, since
they were addressing the non-singlet component, by far the dominant one in
the large $z$ region. We deal instead with the whole
set, including mixings with gluons and antiquarks: a more complete analysis 
of $D$ mesons production in $e^+e^-$ collisions at LEP with the full set,
extending also down to small-$z$ values, will be presented elsewhere
\cite{cgprog}.

The parameter $\langle n_H\rangle$ in eq. (\ref{nonpert}) 
is the mean multiplicity of charmed mesons 
produced in the process, i.e. how many mesons does the non-perturbative
fragmentation of a $c$ quark produce.
In the analysis of Ref. \cite{cn} the normalization condition fixed by 
$\langle n_H\rangle$ has not been determined.
We have therefore used the mean multiplicity 
simulated by the HERWIG generator in the process $e^+ e^-\rightarrow c \bar c$ 
at 90 GeV. This provides us with the last missing parameter for fully defining
the non perturbative part of the fragmentation function. The values we used are
summarized in Table \ref{table1}: 
they describe either the non perturbative fragmentation
of a charm into a $D^0$ or a $D^{*0}$, or alternatively of an anticharm into a 
$\bar D^0$ or a $\bar D^{*0}$. As a double check, the same multiplicities have 
also been extracted from JETSET, by fragmenting a charm quark of 5.3 GeV
energy. They
are also reported in Table \ref{table1}, and can be seen to agree well. It must
be noted that the $D$ mesons multiplicities include the feed-down from the $D^*$
decays. It is also worth noticing that the non-perturbative fragmentation
function (\ref{nonpert}), with the parameters shown in Table \ref{table1},
compares nicely with a Peterson fragmentation function with $\epsilon =
0.06$, used for instance in \cite{fmnr} to describe the transition from
$c$ quarks to $D$ mesons.

\begin{table}
\begin{center}
\begin{tabular}{|c|c|c|c|c|}
\hline
\hline
Meson & $\alpha$ & $\beta$ & $\langle n_H \rangle$ & $\langle n_H \rangle$ \\
\hline
$D^0$ & 1.0 & 3.67 & 0.58  & 0.58 \\ 
$D^{*0}$ & 0.6 & 5.4 & 0.3 & 0.29 \\ 
\hline
& \multicolumn{2}{|c|}{Colangelo/Nason} & HERWIG & JETSET \\
\hline
\end{tabular}
\end{center}
\caption{\label{table1}\small Collection of parameters which describe the
non-perturbative part of the fragmentation functions. In the Colangelo-Nason
paper, ref. \protect\cite{cn}, these $\alpha$ and $\beta$ where obtained with
$\Lambda_5$ = 100 MeV.}
\end{table}

\begin{figure}[t]
\begin{center}
\epsfig{file=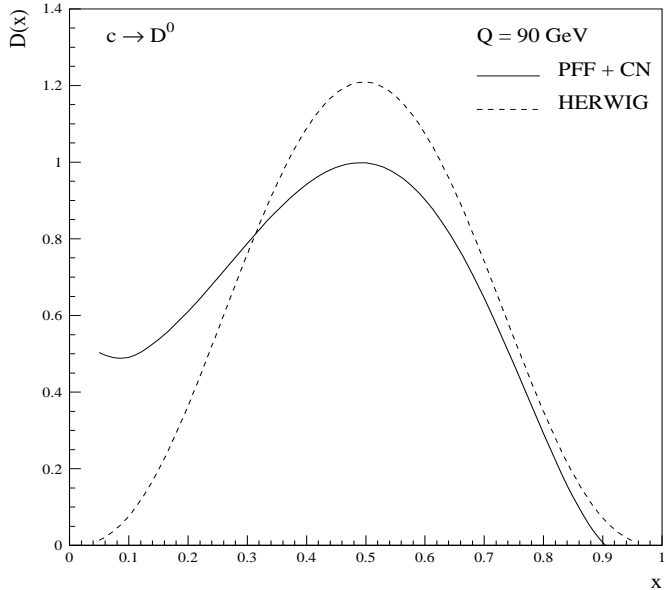,
              bbllx=30pt,bblly=160pt,bburx=540pt,bbury=660pt,
             width=9.cm,height=8cm,clip=}
\parbox{12cm}{
\caption{\label{fig1}\small Comparison between the $D_c^{D^0}$ fragmentation
function produced by HERWIG at 90 GeV and the one given by our ansatz
(\protect\ref{ansatz}) with the non-perturbative parameters by Colangelo and
Nason (see table \protect\ref{table1}).
}
}
\end{center}
\end{figure}

The use of an explicit parametrization for the FF at a given scale 
extracted from HERWIG has been 
successfully  made in the past for the case of the inclusive production 
of light and strange mesons \cite{simona}. In these papers not only the
multiplicity but also parameters $\alpha$ and $\beta$  dictating the shape of
the FF (see Table \ref{table2}) were
extracted from HERWIG at some large scale. 
We have refrained from doing so in this work, since the large mass of the charm
quark provides us with a solid ground for evaluating in pQCD at least the
perturbative part of the fragmentation function, see eq. (\ref{ansatz}). For the
sake of completeness we do however provide in Fig. \ref{fig1} a comparison 
between the
fragmentation function which we obtain by evolving the ansatz of eq.
(\ref{ansatz}) and that we get using the HERWIG parametrization given in Table
\ref{table2}, at a scale of 90 GeV. The value of $\Lambda_5$ = 100 MeV has been
used in the evolution, for consistency with the Colangelo-Nason fits \cite{cn}. 
The Sudakov form factor \cite{melenason} has also been
included, as it was taken into account in ref. \cite{cn}.
Its effects are however small, of the order of a few percent, both on the
fragmentation function itself and on the cross sections which will follow.

A number of comments about this plot are in
order. The discrepancy in the small $x$ region is due to the PFF being enhanced
by mixing with the gluon splitting kernel, while the HERWIG result is
suppressed by phase space constraints. This difference is however of no
practical importance in the evaluation of hadronic cross sections, since such
small $x$ values probe the very large $p_T$ tail of the kernel cross section,
where the latter is small. The discrepancy around the maximum can instead be
considered as a normalization off-set, which could be eliminated by a
fine tuning of both FFs to some experimental data. It is however worth noticing
that the present accuracy of the data is not better than the uncertainty
originating from the difference of the two FFs. Moreover,
the consequences on the observable hadronic cross section are small.

\begin{table}
\begin{center}
\begin{tabular}{|c|c|c|c|c|}
\hline
\hline
     & $\alpha$ & $\beta$ & $N^i_H$ & $\langle n^i_H \rangle$ \\
\hline
$c \to D^0$ & 2.75 $\pm$ 0.08 & 2.72 $\pm$ 0.03 & 53.6  & 0.58 \\ 
$g \to D^0$ & 1.12 $\pm$ 0.51 & 0.36 $\pm$ 0.19 & 0.0045  & 0.0013 \\ 
\hline
\hline
$c \to D^{*0}$ & 2.01 $\pm$ 0.06 & 2.42 $\pm$ 0.08 & 12.4 & 0.30 \\ 
$g \to D^{*0}$ & 0.16 $\pm$ 0.08 & 0.016 $\pm$ 0.03 & 0.008 & 0.007 \\ 
\hline
\end{tabular}
\end{center}
\caption{\label{table2}\small Collection of parameters which describe the
fragmentation functions produced by HERWIG at 90 GeV, parametrized as $D_i^H(z)
= N^i_H (1-z)^{\alpha_i} z^{\beta_i}$. It also holds $\langle n \rangle = N
B(\alpha+1,\beta+1)$, $B$ being the Euler Beta function representing the
normalization integral of the FF. Note that the parameters $\alpha$ and $\beta$
do not have to coincide with those of eq. (\protect\ref{nonpert}), because they
include also the perturbative component of the FF.}
\end{table}

\section{Results}

\begin{figure}[t]
\begin{center}
\epsfig{file=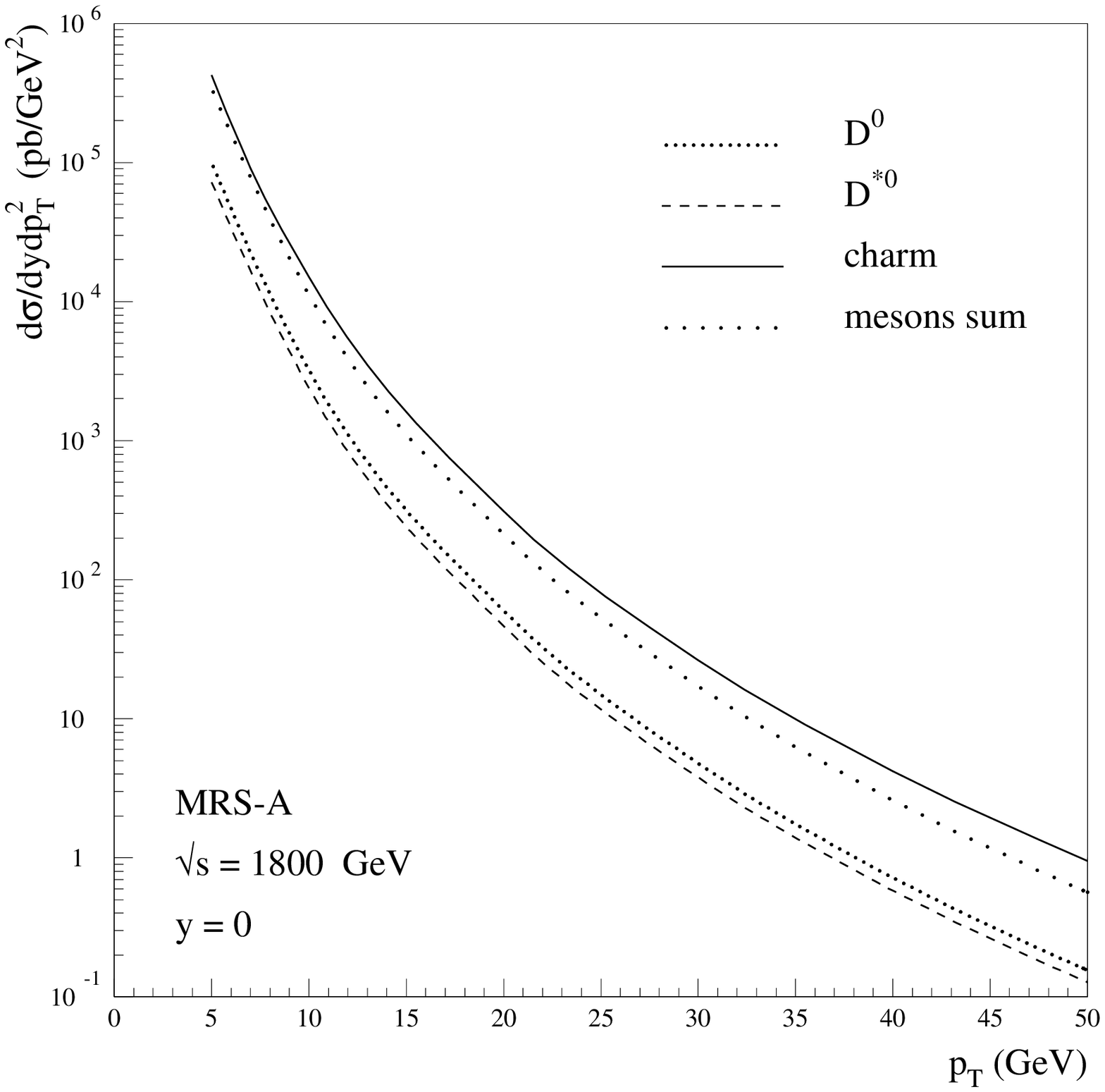,
              bbllx=30pt,bblly=160pt,bburx=540pt,bbury=660pt,
             width=7.cm,clip=}
\hspace{.5cm}
\epsfig{file=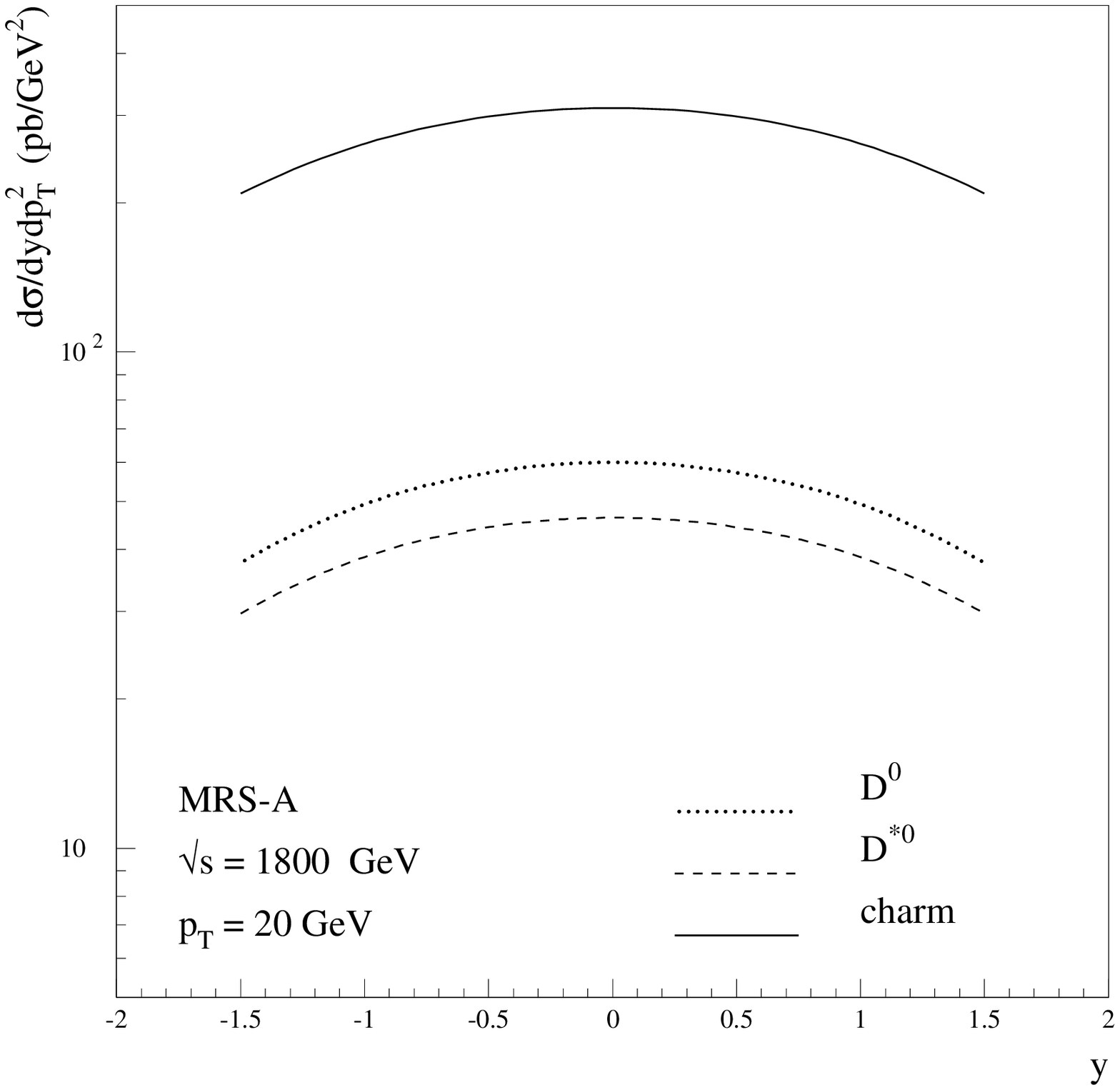,
             bbllx=30pt,bblly=160pt,bburx=540pt,bbury=660pt,
             width=7.cm,clip=}
\parbox{12cm}{
\caption{\label{fig2}\small Cross sections for mesons production at the
Tevatron, as predicted by our fragmentation functions approach. Comparison
with the pure charm quark cross section is also shown. The mesons sum also
includes $D^+$ and $D^{*+}$ contributions.
}
}
\end{center}
\end{figure}

Using the perturbative initial conditions (\ref{DQQ},\ref{DgQ},\ref{DqQ})
and the non-perturbative 
parametrization (\ref{nonpert}) with the parameters summarized in Table 
\ref{table1}, we have a NLO evaluation of the parton 
FF in $D,D^{*}$ mesons at a scale $\mu_0 \approx m_c$.
Then using the DGLAP evolution
equations to NLO accuracy one can evaluate the fragmentation functions set
at any desired factorization scale $\mu$. By convoluting this set with the
NLO kernel cross sections for massless 
parton scattering \cite{aversa} one gets a
prediction for $D$ and $D^*$ production at large $p_T$ at a hadron collider.
Figures \ref{fig2}a,b show the $p_T$ and rapidity spectra for the Tevatron.
Also shown (solid line) is the pure charm quark NLO cross section, 
as predicted by the
Perturbative Fragmentation Function approach. These results have been obtained
with the MRS-A structure functions set \cite{mrsa} and with a $\Lambda_5$
value of 100 MeV. The Sudakov form factor \cite{melenason} has also been 
included: we notice here once more that its effects are numerically small, 
below 10\%, in the $p_T$-rapidity range we have considered.

As expected the $D$ mesons cross
sections lie below the charm quark one. It can however be checked that summing
them up and introducing an additional factor of two to allow for $D^+$ and
$D^{*+}$ state (which are assumed to fragment like the neutral ones) the charm
cross section is almost reproduced (wide-dotted line). The small residual gap
is given by the mesons FF being softer and therefore producing a smaller
cross section.

Many uncertainties do of course affect this result, though not shown in the
plots. First of all,
the mesons fragmentation functions will share all the uncertainties related to
the heavy quark PFFs on which they are built on. 
Factorization scale and initial scale dependencies, of the kind studied in
\cite{cg1}, will also appear here with a similar behaviour, leading to an
uncertainty of order 20-30\%. These fragmentation
functions will also share the same shortcomings of the PFFs. This means that
their description is not accurate at low $p_T$ and at the edges of phase space
(see also \cite{cg2} for a discussion of this point),
where unresummed higher order corrections and non-perturbative effects play an
important role. This leads to the impossibility of performing a meaningful
comparison with the $D^*$ photoproduction data collected 
at HERA, since the minimum $p_T$, of the order of 2-3 GeV, is too low and the
edges of phase space in rapidity are probed.

A second kind of uncertainty is related to the determination of the
non-perturbative parameters $\alpha$, $\beta$ and $\langle n_H \rangle$. The
set we have chosen was fitted to ARGUS data, and has been picked mainly for
illustrative purposes, although, as stated above, it reproduces fairly well the
actual OPAL data from LEP. An analysis of all LEP data, when available,
will certainly lead to a more
precise determination of these parameters. Of course also a detailed
measurement of the $D$, $D^*$ inclusive cross sections at the Tevatron would be
very helpful and give complementary information on the FF, particularly for the
gluon terms.

To conclude, we have presented a model for the $D$ and $D^*$ fragmentation
functions based on PFF for heavy quarks complemented by a factorized
non-perturbative term describing the quark-meson transition.
Predictions have been given for large $p_T$ charmed mesons production at the
Tevatron.

\vspace{.5cm}
\noindent
{\bf Acknowledgements.} S.R.'s work supported in part by the ``Maria Rossi'' 
fellowship from Collegio Ghislieri of Pavia,
through the U.S. Department of Energy under contract DE-AC03-76SF00098.

\newcommand{\zp}[3]{Z.\ Phys.\ {\bf C#1} (19#2) #3}
\newcommand{\pl}[3]{Phys.\ Lett.\ {\bf B#1} (19#2) #3}
\newcommand{\plold}[3]{Phys.\ Lett.\ {\bf #1B} (19#2) #3}
\newcommand{\np}[3]{Nucl.\ Phys.\ {\bf B#1} (19#2) #3}
\newcommand{\prd}[3]{Phys.\ Rev.\ {\bf D#1} (19#2) #3}
\newcommand{\prl}[3]{Phys.\ Rev.\ Lett.\ {\bf #1} (19#2) #3}
\newcommand{\prep}[3]{Phys.\ Rep.\ {\bf C#1} (19#2) #3}
\newcommand{\niam}[3]{Nucl.\ Instr.\ and Meth.\ {\bf #1} (19#2) #3}
\newcommand{\mpl}[3]{Mod.\ Phys.\ Lett.\ {\bf A#1} (19#2) #3}

\end{document}